# Doppler dual-comb coherent Raman spectromicroscopy

Florian M. Schweizer[1], Hannah Terrasa[1], Manish Garg[1,*]

[1]Max Planck Institute for Solid State Research, Heisenbergstr. 1, 70569 Stuttgart, Germany

*Corresponding author. Email: mgarg@fkf.mpg.de

**Chemical imaging enabled by Raman processes is crucial to investigating biological and chemical samples in a label-free manner. Stimulated Raman spectroscopy (SRS) overcomes the key limitation associated with low signal levels in spontaneous Raman spectroscopy, however, at the expense of probing only narrow Raman bands. Time-domain implementation of coherent anti-Stokes Raman spectroscopy (CARS) by dual frequency combs can achieve broad Raman bandwidths; nevertheless, its execution is demanding due to strenuous temporal-synchronization of two independent ultrashort laser sources. Here, we introduce time-domain coherent Raman spectroscopy utilizing two frequency combs generated by the Doppler effect from a single ultra-broadband laser source. In contrast to CARS, in our approach, the interference of impulsively launched vibrations by two broadband frequency combs ($\tau \sim 6$ fs) periodically modulates the Kerr nonlinear response of the medium, leading to cross-phase modulation (XPM) experienced by both the combs. This phase modulation leads to spectral broadening and periodic modulation in the anti-Stokes region of the combs. Down-conversion by a factor of $\sim 10^{-8}$ in the frequency of the vibrations enabled by the dual-comb approach empowered us to use photon-counting methodology in the anti-Stokes region. This makes our technique extremely versatile, background-free, sensitive and fast (millisecond acquisition times), in probing a range of samples from wide bandgap dielectrics and liquids to individual micro-particles with nondestructive pulse energies ($\sim 100$ pJ) incident on the sample. Owing to the higher-order nonlinearity involved in the XPM process, we achieved $\sim 2.5$ times improvement in diffraction-limited spatial resolution ($\sim 280$ nm) in ultra-broadband chemical imaging of a $\sim 8$ $\mu$m bead of poly-methyl-methacrylate. We expect our approach to be able to capture three-dimensional Raman holographs of single protein molecules.**

**Introduction**

Optical spectroscopy and microscopy have emerged as powerful tools to probe biological and chemical functionalities in the last decades. Fluorescence[1,2] and scattering[3-5] based super-resolution imaging techniques can now routinely achieve spatial resolution significantly below the optical diffraction limit (~ 10 nm); nevertheless, they fail to capture the chemical fingerprint of the molecules. Moreover, it often requires labelling the samples with fluorescent markers. Raman spectromicroscopy[6] can chemically image molecules in a label-free manner[7-9], however, intrinsically low signal levels and large fluorescence backgrounds[10] dramatically limit the applicability of this approach. Coherent Raman techniques, e.g., SRS overcome these limitations, but can only probe narrow Raman bands at a time and achieve only diffraction limited spatial resolution[11-18]. On the other hand, time-domain implementation of CARS[19,20] by dual frequency-comb approach[21,22] can achieve ultra-broad Raman bandwidth and fast acquisition times (~ 200 $\mu$s), however, the temporal synchronization of two ultrashort laser sources with few fs timing jitter continues to be challenging[23]. Apart from the coherent Raman features, large non-resonant backgrounds (NRB) often conceal the CARS signal[24,25].

In the current work, we introduce a time-domain, background-free, coherent Raman spectroscopic approach by employing dual frequency combs generated from a single broadband laser source ($\tau$ ~ 6 fs) via the Doppler effect[26]. Reflection off the surface of a mirror moving at an ultrasonic frequency (~ 20 kHz) shifts the underlying combs lines of an ultrashort laser pulse by ~ ± 6Hz. Interference of the impulsively launched vibrations by two orthogonally polarized broadband frequency combs, with one of them being Doppler frequency shifted, periodically modulates the Kerr nonlinearity ($n_2$) of the molecules[27-29], resulting into cross-phase modulation[27,30] experienced by both the combs. Polarization selective detection of this phase modulation results into periodic variation of the spectral broadening induced in the anti-Stokes region of the combs. The down-conversion of the vibrational frequencies from ~100 THz to the MHz region enabled by the dual-comb methodology[21,31,32] empowers us to use relatively slower single-photon detectors in the anti-Stokes region to time-resolve the vibrations. Fast acquisition times of ~10 milliseconds were achieved in ultra-broadband Raman measurements (~ 200 cm$^{-1}$ to 3200 cm$^{-1}$), which can be made even faster with further development. Photon counting makes our technique extremely sensitive to detect vibrations in a range of samples from bulk dielectrics (crystalline quartz) and liquids (dimethyl sulfoxide (DMSO) and toluene) to isolated micro particles of poly-methyl-methacrylate (PMMA). This phase-sensitive XPM approach to record vibrations eliminates unwanted non-resonant background and fluorescence encountered in CARS[33,34].

In contrast to SRS and CARS, the higher order nonlinearity associated with the XPM process enables superior spatial resolution (~ 280 nm) compared to the diffraction limited spatial resolution (~ 700 nm) in broadband chemical imaging of a ~ 8 μm PMMA micro-particle, which could be improved to ~ 100 nm by utilizing higher numerical aperture objectives and beam shaping techniques. Furthermore, spectral-domain measurement of the XPM signal shows broad anti-Stokes spectra, thus paving the way towards coherent Raman holography of single protein molecules, which has been inaccessible until now.

**Ultra-broadband Doppler dual frequency combs**

In our experiments, ultrashort laser pulses from a broadband Ti:Sapphire oscillator (~ 700-1050 nm, $\tau$ ~ 6 fs and repetition rate ~ 80 MHz) were separated into two optical paths (Fig. 1a). In one optical path, the pulses were reflected off a mirror oscillating at an ultrasonic frequency[35,36] (~ 20 kHz, amplitude ~ ± 90 μm) and underwent a Doppler shift in the underlying frequency comb lines[26] with respect to the fundamental comb lines in the other optical path, hereafter, referred to as the 'pump' pulses (Fig. 1a). The Doppler shifted comb lines constitute the 'probe' pulse. The $n^{th}$ comb-line in the probe pulses is frequency shifted to ~ $nf_r$ $(1 \pm \frac{v}{c})$, where $f_r$ is the repetition rate of the laser pulses, $v$ is the speed of the ultrasonic mirror and $c$ is the speed of light. Since the speed of the ultrasonic mirror is non-uniform, i.e. maximum at the zero-crossing, whereas minimal at the turning points, the Doppler frequency shift is time varying. Thus, the time-delay introduced by the ultrasonic mirror between the pump and probe pulses is also non-uniform, and was calibrated by a reference Michelson interferometer with a continuous wave HeNe laser (Fig. 1a). The variation in the intensity of the interference fringes as a function of the movement of the ultrasonic mirror as recorded by a fast photodiode is shown in Fig. 1b. The interference fringes become narrowly separated around the zero-crossings of the mirror, whereas are very broad near the turning points[35,36]. The blue curve in Fig. 1b and the associated black rectangles indicate the motion of the ultrasonic mirror; at the maxima of the blue curve, the mirror is at the turning point. The delay-axis in the measurement was made uniformly varying (see methods for details) by calibrating the position of the maxima of the HeNe interference fringes to the wave-cycle of the laser ($\lambda_{HeNe}$ ~ 632.8 nm, wave-cycle ~ 2.1 fs).

The underlying frequency comb spectra of the pump and probe pulses is shown in Fig. 1c, where we have ignored the carrier-envelope offset frequency of the comb lines, as its effect is eventually cancelled out. The maximum frequency difference between the individual comb lines of the probe pulses will be ~ $f_r \pm \delta f_r$ ($f_r$ ± 6 Hz), whereas the comb lines in the pump pulse will be separated by $f_r$. This is akin to the dual comb methodology employing two independent laser sources of slightly different repetition rates. As a first illustration of the capability of the Doppler dual frequency combs, we measured the second-harmonic generation based fringe-resolved autocorrelation between the pump and probe pulses with an approximately 100 μm thick Beta-Barium Borate (BBO) crystal. The slight difference in the second harmonic of the $n^{th}$ comb line in both the frequency combs will down-convert the carrier frequencies of the laser pulses by ~

$f_c \frac{\delta f_r}{f_r}$, where $f_c$ is the central carrier frequency of the laser pulses (~ 400 THz). This consequently enables measurement of the pulse autocorrelation on an only ~ 25 μs timescale with a fast photodiode (response time ~ 1 ns) as shown in Fig. 1d. The down-conversion of the carrier frequency of the laser pulses due to the Doppler effect is explained in the methods section. Because of the narrower spectral phase matching bandwidth of the used BBO crystal, the second order autocorrelation of the laser pulses appears slightly broader than the duration of the laser pulses.

**Molecular vibration mediated modulation of Kerr nonlinearity**

On interaction with molecules, both the pump and probe pulses can impulsively excite vibrations[21], e.g., by an upward transition from the n[th] comb line ensued by a downward transition from the m[th] comb line (Fig. 1c). The pump and probe pulses are orthogonally polarized. The vibrations launched by the two pulses will have a slight offset in their frequencies: $f_{vib} = (m - n) f_r$ and $f'_{vib} = (m - n)(f_r + \delta f_r)$. The slight difference in the vibrational frequencies will lead to their beating at a slower frequency of $f_{vib} \frac{\delta f_r}{f_r}$, thus effectively resulting in the down conversion of the frequencies by a factor of nearly $10^{-8}$ (Fig. 1e), e.g., from ~ 100 THz to ~ 1 MHz. This down-conversion in the frequency of the molecular vibrations is explained in the methods section.

The launched vibrations in the medium by the pump and the probe pulses periodically modulate the Kerr nonlinear response[27,29], i.e., inducing a time-varying refractive index, eventually leading to XPM experienced by both the pulses[30]; $\Delta\phi(t) = \frac{2\pi}{\lambda} n_2 \left( I_{probe}(t + \tau) + \kappa\, I_{pump}(t) \right) L$, where $\tau$ is the delay between the pulses, $\lambda$ is the wavelength of the laser pulses, $I$ is the intensities of the pulses, $\kappa$ is a coefficient often close to 2 for most materials[37] and $L$ indicates the interaction length in the sample. XPM periodically modulates the intensity of the spectral broadening in the anti-Stokes region (< 700 nm) of the two frequency combs, $\omega(t) = -\frac{\partial \Delta\phi(t)}{\partial t}$. In absence of the pump pulses, probe pulses alone can result in spectral broadening in the anti-Stokes region by the $\chi^{(3)}$ process (Fig. 1e), a phenomenon often referred to as self-phase modulation[38] (SPM). The mechanism of modulation of the XPM signal by the molecular vibrations will be discussed later in the text. In the experiments, only the anti-Stokes emission arising from the XPM experienced by the probe pulses was recorded, by rotating the polarizer in front of the single-photon detector along the polarization axis of the probe pulse (Fig. 1a). An additional non-resonant photon interaction from the probe pulse with the vibrationally excited molecule (Fig. 1e) leads to the measured anti-Stokes emission (downward blue arrow in Fig. 1e). In contrast to CARS, spectral background resulting from the non-resonant four-wave-mixing processes[24,34] (NRB) and fluorescence is dramatically reduced in the anti-Stokes spectral region in this adaptation of coherent Raman spectroscopy.

The down-conversion of the vibrational frequencies to the MHz region permitted the use of relatively much slower photon-counting methodology[39] (~ 20 MHz) in the anti-Stokes spectral region to track molecular vibrations in real-time. Left-panels in Fig. 2 show the variation of the XPM signal measured as a function of the delay between the pump and the probe pulses for dimethyl sulfoxide (DMSO, Fig. 2a), toluene (Fig. 2b) and crystalline z-cut Quartz of ~ 50 μm thickness (Fig. 2c). Right-panels in Fig. 2 show the corresponding fast Fourier transformation (FFT) of the time-resolved XPM traces. Ultra-broadband Raman spectra were captured for all the samples, ranging from the $A_{1b}$ mode of quartz at ~ 462 cm$^{-1}$ to the C−H stretching modes in DMSO and toluene at ~ 3000 cm$^{-1}$. One delay scan of ~ 1.2 ps by the ultrasonic mirror requires approximately 25 μs. Forward and backward movements of the mirror are symmetric (Fig. 1b), therefore the signals from these scans were averaged together. An acquisition time of one second was used for the time-resolved traces shown in Fig. 2, which corresponds to averaging over ~ $40 \times 10^3$ pump-probe delay scans. The impact of averaging the pump-probe delay scans on the photon-counting methodology can be seen in Fig. 3. An acquisition time of 10 ms (corresponding to ~ 400 delay scans) in the measurement on DMSO shows a reasonable FFT spectrum, nevertheless, with lower signal to noise levels, which is undoubtedly much better for longer averaging times. The fast acquisition times, in the millisecond range, makes the current XPM technique nearly hundred times faster than spontaneous Raman spectroscopy[10]. Use of faster single-photon detectors in the future can lower the acquisition times even further.

The sensitivity of the current XPM technique to track the vibrations can be recognized from the measurement on the Quartz sample. Time-resolving phonons in wide bandgap (~ 9 eV) Quartz has previously required using several orders of magnitude higher pulse energies, ~ 10 μJ (or higher), compared to ~ 100 pJ used in the current work, either by high-harmonic spectroscopy[40] or by transient reflection[41].

**Multi-wave mixing detection of molecular vibrations**

To decipher the mechanism underlying the detection of coherent molecular vibrations in the experiments, we performed fluence dependence measurements in a liquid and a solid sample, DMSO and Quartz. The variation of the FFT amplitude of the vibration corresponding to the C−H stretching mode in DMSO (~ 2913 cm$^{-1}$) as a function of the increasing fluence of the pump and probe pulses is shown in Fig. 4a and Fig. 4b, respectively. The power of the probe and pump pulses in Fig. 4a and Fig. 4b was kept constant at 29 mW and 19 mW, respectively. A quadratic scaling was measured for the variation of the fluence of the pump pulses, whereas a cubic scaling was measured for the variation of the fluence of the probe pulses. Similar fluence dependence on the pump and probe pulses was measured for the $A_{1b}$ vibrational mode (~ 462 cm$^{-1}$) of the Quartz sample (Fig. 4c and Fig. 4d).

A quadratic dependence of the vibrational amplitude on the pump fluence rationalizes the impulsive excitation of the vibrations by a two-photon process (Fig. 1e). A cubic dependence of the vibrational amplitude observed for the probe fluence implies an intra-pulse four-wave mixing ($\chi^{(3)}$) process. Spectral-domain measurement of the XPM process shown in Fig. S1 of the supplementary materials reveals that the dominant contribution of the XPM signal comes close to ~ 670-690 nm, which is the anti-Stokes spectral region of the probe pulse. This precludes the contribution of the third-harmonic generation from the probe pulse in the XPM signal, involving three consecutive upward photon transitions, yielding to anti-Stokes emission at ~ 300 nm. The observed cubic dependence and the measured spectral bandwidth of the XPM signal connotes that the dominant contribution comes from the photon processes depicted in Fig. 1e. Since the pump and probe pulses are identical in their characteristics, except their polarizations, it is reasonable that probe pulses also impulsively excite the vibrations as the pump pulses. However, an additional photon interaction from the probe pulses with the vibrationally excited molecules leads to the anti-Stokes emission (Fig. 1e), consequently leading to the measured cubic fluence dependence. Identical fluence dependence observed for a liquid and a solid sample infers that the process of modulation of the Kerr nonlinear response (XPM) by the impulsively excited vibrations is very general in nature.

The photon transition from the vibrationally excited molecule leading to the anti-Stokes emission (Fig. 1e) can arise from a Raman like process or from a purely electronic process. The ultrashort duration of the laser pulses ($\tau$ ~ 6 fs) used in the current experiments are not commensurate with the Raman response times[38], which will be similar to the oscillation period of the probed vibrations (~ 11 to 80 fs), effectively making the probability of the Raman transitions negligible. Hence, the anti-Stokes emission is dominated primarily by the electronic transitions. Figure S1 in the supplementary materials shows the spectral-domain characterization of the XPM signal measured from a single PMMA bead. Because of the $O-CH_3$ bending mode of PMMA at ~ 812 cm$^{-1}$, the anti-Stokes emission extends down to ~ 665 nm, since the corresponding driving photon (Fig. 1d) for this emission is at the blue spectral edge of the probe pulses (~ 700 nm, Fig. S2). In close analogy, the anti-Stokes emission for the $C-H$ stretching mode of PMMA at ~ 2957 cm$^{-1}$ in the XPM signal extends much further down to ~ 630 nm (Fig. S1). The scattering cross-section of non-resonant electronic transitions are orders of magnitude higher than that of the Raman transitions, therefore making XPM based coherent Raman approach considerably more sensitive than CARS.

**Sub-diffraction limited chemical imaging**

We now demonstrate the hyper-spectral chemical imaging capability of our technique. Single carboxylated PMMA beads of ~ 8 μm size (Fig. S3) were drop-casted on a Quartz wafer and mounted on a two-dimensional piezo stage (Fig. 1a). Such polymer beads have regularly been used to benchmark coherent Raman techniques, owing to their close resemblance with biological cells. Firstly, we perform z-sectioning[42], by acquiring two-dimensional images of the SPM signal, and moving the bead at several positions along the focal axis, as shown in Fig. 5a. The estimated focal spot size of the laser pulses was ~ 1.2 μm. The images show the spatial variation of the SPM signal generated only by the probe pulses and acquired by the single-photon counter. Because of the spherical geometry of the bead, it appears to be smaller in the 2D image as it is moved away from the focus (last panel in Fig. 5a). Whereas at the center of the focus (~ − 0.3 μm), the bead appears bigger in size and exhibits the maximum signal, however, the image reveals an empty structure at the core of the bead. This plausibly arises from the lower distribution of the molecules at the core of the bead compared to its periphery. The magnitude of third-order nonlinear susceptibility, $\chi^{(3)}$, is directly linked to the local electronic density-of-states[43] (DOS). A higher density of molecules implies a higher electronic DOS and correspondingly a higher $\chi^{(3)}$ and SPM signal.

Figure 5b shows the time-resolved XPM signal measured as a function of the delay between the pump and probe pulses with the cross annotated in the second last panel of Fig. 5a indicating the position of the laser focal spot on the bead. The FFT spectrum of the XPM signal (Fig. 5c) reveals multiple Raman peaks linked to the stretching modes of C−H (~ 2957 cm$^{-1}$), C=O (~ 1725 cm$^{-1}$), and C−C (~ 812 cm$^{-1}$), the bending mode of O − CH$_3$ (~ 1451 cm$^{-1}$), and the A$_{1b}$ mode (~ 462 cm$^{-1}$) of the underlying Quartz substrate.

Variation of the amplitude of the Raman peaks retrieved from the time-resolved XPM measurements along the orange line annotated in the second last panel of Fig. 5a is shown in Fig. 5d. The Raman bands are color coded one-to-one in Fig. 5c and Fig. 5d. The spatial distribution of the SPM signal (orange-curve in Fig. 5d) is visibly broader than the spatial distribution of the amplitude of the Raman peaks. The SPM signal generated only by the probe pulses arises from the $\chi^{(3)}$ process, which follows the third order nonlinearity, whereas the Raman peaks retrieved from the XPM signals exhibit a second order nonlinearity on the pump pulse and a third order nonlinearity on the probe pulse (Fig. 4). A higher order nonlinearity in the XPM signal brings the estimated spatial resolution to ~ $\frac{680}{2 \times NA \times \sqrt{6}}$ nm ~ 280 nm, where NA is the numerical aperture of the objectives used in the experiments (~ 0.5). This spatial resolution is significantly better than the diffraction limited resolution of ~ 680 nm. The spatial resolution was estimated by considering the dominant spectral component in the XPM signal as shown in Fig. S1.

The spatial resolution in the current XPM based coherent Raman technique is significantly better than other coherent variations, such as SRS[44] and CARS[45]. Higher-order nonlinear wave-mixing CARS[46] has been shown to achieve spatial resolution of < 200 nm, nonetheless, it can only capture one Raman band. The estimated intensity of the laser pulses on the bead was ~ 10$^{12}$ W/cm$^2$, which in principle can allow higher-order six and eight wave-mixings leading to the XPM signal, however, the $\chi^{(3)}$ contribution (Fig. 1e) dominates all other higher-order wave mixing signals (Fig. 4). Even though the peak intensities of the laser pulses were very high, the ultrashort duration of the pulses and lower pulse energies (< 100 pJ) prohibit the deposition of the thermal energy into the beads, thus ensuring their stability. The used pulse energies are nondestructive to biological cells.

**Concluding remarks**

The capability, demonstrated here, to track molecular vibrations by cross-phase modulation in a dual-comb pump-probe configuration opens the possibilities to perform Raman holography of single protein molecules (e.g. nuclear pore complexes, size > 100 nm) and monitor ultrafast changes in their structures, e.g., during protein folding. The utilized photon counting methodology in the background free anti-Stokes spectral region of the driving laser pulses makes the technique sensitive and generic to track vibrations in a multitude of samples, from broad bandgap dielectrics and liquids to single micro-particles. Ultra-broadband Raman spectra (~ 200 $cm^{-1}$ to ~ 3200 $cm^{-1}$) can be captured with an acquisition time of only ~ 10 milliseconds, with a sub-diffraction limited spatial resolution of ~ 280 nm. The spatial resolution can be brought down below 100 nm by beam shaping techniques and by the use of higher NA reflective objectives. Wide-field (~ 20 μm) adaptation of the current approach paves the way towards real space-time visualization of chemical processes unfolding inside a biological cell and in the liquid phase.

**Methods**

**Experimental Setup**

The ultra-broadband laser pulses ($\lambda \sim 700-1050$ nm and $\tau \sim 6$ fs) from a Ti:Sapphire oscillator (Element 2, Spectra-Physics) were pre-compensated for their linear chirp by a set of chirped dielectric mirrors, with a group-delay dispersion (GDD) for one pair of reflection of $\sim -120$ fs$^2$. The unequal amount of the dispersive material in the beam paths of the pump and probe pulses (Fig. 1a) and the additional glass of the cuvette for the liquid samples was compensated by inserting an appropriately thick fused silica window in the two optical paths.

For the ultrasonic mirror, an $\sim 1$ mm thick silver mirror (PFR10-P01, Thorlabs) was cut and glued with epoxy (EPO-TEK® H77, Meridian) directly onto a Sonotrode (UIP500hdT with the BS4d18 and B4-3.0 for boosting the amplitude, Hielscher Ultrasonics). The movement of the ultrasonic mirror was monitored by the interference of the reference HeNe laser (HRS015B, Thorlabs). For lower thermal drifts during operation, the Sonotrode was locally air-cooled.

The ultra-broadband laser pulses were focused onto the sample through a reflective objective (36×, NA 0.5, 50102-02, Newport). The sample was mounted on a stack of two piezo stages (Physik Instrument) mounted on a 3D-stage (Newport) for a bigger movement range. The light emanating from the sample was collected in the forward direction by another objective (50×, NA 0.55, Mitutoyo). The PMMA bead sample was imaged with a camera (CS165CU, Thorlabs). The anti-Stokes signal from the samples was detected with a single-photon detector (SPCM-AQRH-14, PerkinElmer). The signal was acquired directly with an oscilloscope (MSO7054A, Keysight). All experiments were performed at ambient conditions.

**Down-conversion of the frequency of molecular vibrations**

In the following, we demonstrate how the vibrational frequencies from $\sim 100$ THz (3300 cm$^{-1}$) are down-converted to the MHz range, which is slow enough for single photon counters, by employing the Doppler dual-comb methodology. The electric field of the pump laser pulses can be expressed as a summation over the time evolution of the underlying comb lines:

$$E_{Pump}(t) = \sum_n \exp(inf_r t) + c.c., \qquad (1)$$

where $f_r$ is the separation between the individual comb lines. The electric field of the Doppler frequency shifted probe pulses can be expressed as:

$$E_{Probe}(t-\tau) = \sum_n \exp\big(in(f_r + \delta f_r(\tau))(t-\tau)\big) + c.c., \qquad (2)$$

where $\delta f_r(\tau)$ is the Doppler frequency shift imparted by the ultrasonic mirror, which is dependent on the position ($\tau$) of the mirror. $\delta f_r(\tau) = \frac{v(\tau)}{c} f_r$, where $v(\tau)$ is the speed of the ultrasonic mirror. Here, we have ignored the carrier-envelope offset frequency ($f_0$) of the comb lines, as its effect is eventually cancelled out, as well as the respective amplitudes of the comb lines. The electric fields of the impulsively excited vibrations (Fig. 1e) by the two pulses can be expressed as:

$E_{vib}(t) = \sum_{m-n} \exp(i(m-n)f_r t) + c.c.$   and   $E_{vib}'(t-\tau) = \sum_{m-n} \exp(i(m-n)(f_r + \delta f_r(\tau))(t-\tau)) + c.c.$ (3)

Assuming a summation over all the comb pairs ($m$, $n$) exciting the vibrational frequencies, $(m-n)f_r = f_{vib}$ and $(m-n)(f_r + \delta f_r(\tau)) = f_{vib} + \delta f_{vib}(\tau)$.

$E_{vib}(t) \sim \exp(if_{vib}t) + c.c.$ and $E_{vib}'(t-\tau) \sim \exp(i(f_{vib} + \delta f_{vib}(\tau))(t-\tau)) + c.c.$ (4)

The interference of these two vibrational frequencies in the medium will evolve in time as;

$$E_{vib}(t)^* E_{vib}'(t-\tau) \sim \exp\big(i(\delta f_{vib}(\tau)t - \delta f_{vib}(\tau)\tau - f_{vib}\tau)\big) + c.c.$$ (5)

This interference down-converts the frequency of the vibrations from $f_{vib}$ to $\delta f_{vib}(\tau)$

$$\delta f_{vib}(\tau) = (m-n)\big(f_r + \delta f_r(\tau)\big) - (m-n)f_r = (m-n)\delta f_r(\tau) = f_{vib}\frac{\delta f_r(\tau)}{f_r}$$ (6)

The down-conversion factor is $\frac{\delta f_r(\tau)}{f_r} = \frac{\delta f_{vib}(\tau)}{f_{vib}} = \frac{v(\tau)}{c}$.

The delay imparted by the ultrasonic mirror is $\sim 360$ μm in $\sim 25$ μs, giving a mean speed of the mirror of $\sim 14$ m/s, which leads to a down-conversion factor of $\sim 5 \cdot 10^{-8}$. Thus, the oscillation period of $\sim 10$ fs for a vibrational frequency of $\sim 3200$ cm$^{-1}$ is down-converted to $\sim 200$ ns ($\sim 5$ MHz), which is easily resolvable by single photon counters.

It is worth mentioning that the delay imparted by the ultrasonic mirror is non-uniform in nature, thus making the down-conversion factor also delay dependent. The down-conversion factor is $\frac{\pi}{2}$ times worse at the zero-crossings than the mean value, but is getting much better near the turning points.

In close analogy, the changes occurring in the second-harmonic fringe-resolved autocorrelation measurement (Fig. 1d) in timespans as small as $\sim 0.1$ fs could be resolved owing to its down-conversion to $\sim 2$ ns, which is within the response time of the photodiode ($\sim 1$ ns). By shifting the zero-delay of the measurement closer to the turning point, the time variation of the fringes can be resolved even better.

If the delay between the pump and probe pulses would be varied much slowly (~ ms), as in the conventional pump-probe measurements, the single-photon counting approach can still be used, without the need of any down-conversion. Here, the frequency of the vibrations cannot be directly resolved, but its effect on the modulation of the anti-Stokes emission can still be measured by the delay-dependent phase term, $f_{vib}\tau$.

**Sample Preparation**

PMMA microbead sample: An approximately 8 µm sized carboxylated PMMA microbead solution from PolyAn was sonicated for 5 minutes and was further diluted with deionized water using a dilution ratio of 1:10. The resulting solution was sonicated for 5 minutes and drop-casted onto a ~ 50 µm thick z-cut Quartz wafer. The sample was left to dry overnight before its use in the experiments.

**Retrieving delay imparted by the motion of the ultrasonic mirror**

The motion of the ultrasonic mirror was modeled by a sinusoidal function, $x(t) = A(t)sin(\omega_{Sono}t + \varphi)$, where $A$ is the amplitude of the motion, $\omega_{Sono}$ is the frequency of the oscillation and $\varphi$ is the phase. The frequency and the phase of the ultrasonic motion were extracted by finding the turning points of the oscillatory movement from the recorded spectral interferences of the HeNe laser [35] (Fig. 1b). The turning points refer to the maximum and the minimum delays imparted by the ultrasonic mirror; where the direction of the movement of the mirror also reverses. The amplitude of the motion of the ultrasonic mirror ($A$) was calculated by counting the number of interference fringes of the reference HeNe laser between two consecutive turning points. With the amplitude, frequency and the phase, all necessary parameters for the movement of the ultrasonic mirror were known. Thus, for every signal recorded from the single-photon detector on the time-axis, a corresponding delay value between the pump and the probe pulses could be assigned. The signal values were then sorted based on their respective time delays and interpolated to have a uniformly spaced pump-probe delay trace. The used oscilloscope for data acquisition could only capture pump-probe delay sweeps occurring within 10 ms at a time. For longer acquisition times, the zero-delay peaks of the pump-probe traces were aligned, averaged, and Fourier transformed.

**Data availability**

The data that supports the findings of this study are available from the corresponding authors on request.

**Code availability**

The details needed to reproduce the computations have been provided in the "Methods" section and Supplementary Information file.

**Acknowledgments**

We thank Wolfgang Stiepany and Marko Memmler for technical support, Sayooj Sateesh for taking the high magnification optical images of the PMMA beads.

**Contributions**

F.S., H.T., and M.G. built the experimental setup, performed the experiments and analyzed the experimental data. M.G. conceived the project and designed the experiments. All authors interpreted the results and contributed to the preparation of the manuscript.

**Competing interests**

The authors declare no competing interests.

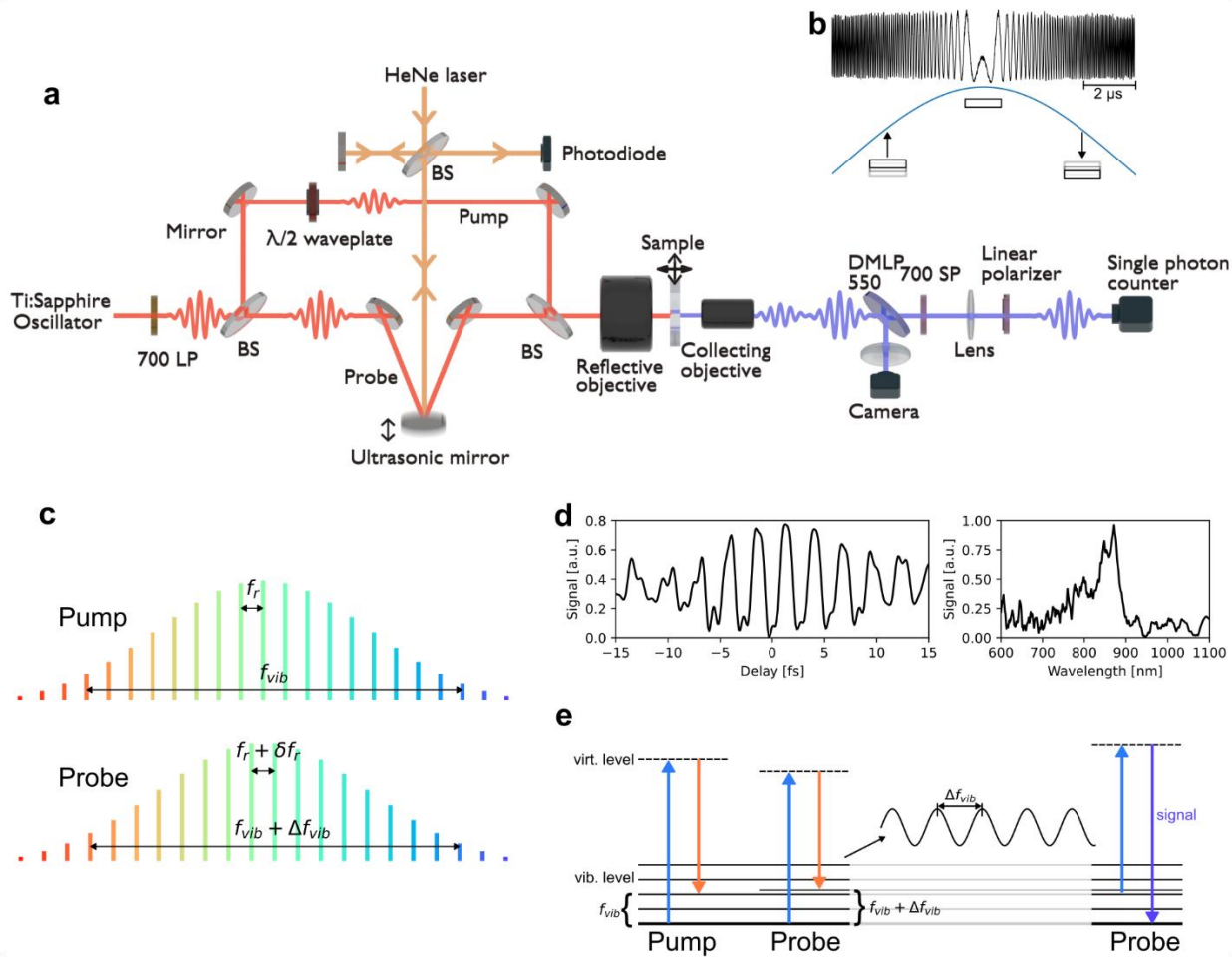

**Figure 1 | Doppler dual-comb coherent Raman spectroscopy. a,** Schematic illustration of the experimental set-up. Ultrashort laser pulses from an ultra-broadband Ti:Sapphire oscillator were split into two optical paths constituting the pump and the probe pulses ($\tau \sim 6$ fs and $\lambda \sim 700 - 1050$ nm). The probe pulses were reflected off the surface of a mirror oscillating at an ultrasonic frequency of $\sim 20$ kHz, before being recombined with the pump pulses and were subsequently focused by a 36× reflective objective (NA $\sim 0.5$) onto the sample mounted on a two-dimensional piezo stage. The anti-Stokes emission emanating from the sample was collected by a 50× objective (NA $\sim 0.55$) and spectrally filtered below 700 nm by a short-pass filter before being focused onto a single-photon detector. The pump and probe pulses were orthogonally polarized in the cross-phase modulation experiments. XPM signal originating from the probe pulses was selectively measured in the experiments by rotating the polarizer in front of the single-photon detector along the polarization axis of the probe pulses. LP: Long-pass filter, BS: Beam-splitter, DMLP: Dichroic long pass filter and SP: Short-pass filter. **b,** Variation of the intensity of the interference fringes of the reference CW beam of HeNe laser measured with a Michelson interferometer (orange arrows in **a**) as a function of the movement of the ultrasonic mirror, scale bar indicates a time span of 2 µs. The delay

introduced by the ultrasonic mirror is non-uniform due to the varying speed of the mirror; maximum at the zero crossings, whereas minimal at the turning points. The maxima of the blue curve indicate the maximum stretch of the mirror, where its speed is nearly zero. **c,** Underlying frequency combs of pump and probe pulses. The comb lines of the pump pulses are separated by $f_r$, which is the repetition rate of the laser source, whereas the comb lines of the probe pulses are separated by $f_r \pm \delta f_r$, $\delta f_r$ is the time-varying Doppler shift due to the non-uniform movement of the ultrasonic mirror. **d,** Left-panel: A second-harmonic generation based fringe-resolved autocorrelation measured between the pump and probe pulses with a ∼ 100 μm thick BBO crystal. Right-panel: Spectrum obtained by FFT of the fringe-resolved autocorrelation trace. Pump and probe pulses were polarized along the same axis for this measurement and an acquisition time of ∼ 25 μs was used. **e,** Schematic depiction of impulsive excitation of vibrations in the sample by a two-photon transition, an upward transition from the blue tail ensued by a downward transition from the red tail of the frequency combs (**c**). Interference between molecular vibrations of slightly different frequencies ($f_{vib}$ and $f_{vib} + \Delta f_{vib}$) launched by the two frequency combs of the pump and probe beams enables down-conversion of their frequencies to the MHz range. An additional photon interaction from the probe pulse with the vibrationally excited molecule leads to the anti-Stokes emission.

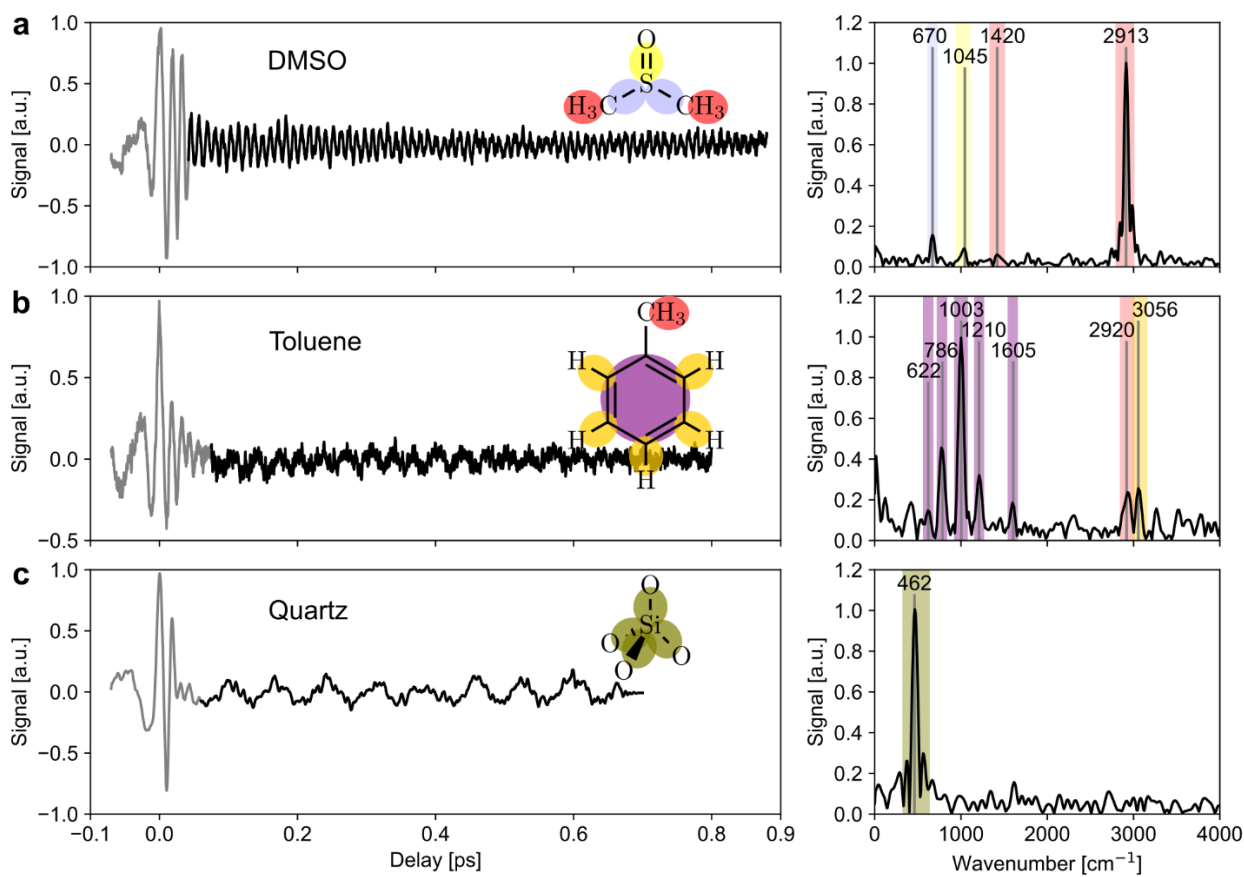

**Figure 2 | Molecular vibrations probed by cross-phase modulation. a, b, c,** Left-panels: Time-resolved XPM signal measured as a function of the delay between the pump and the probe pulses in DMSO, toluene and z-cut Quartz of ~ 50 μm thickness, respectively. Right-panels: Corresponding fast-Fourier transformation spectra of the time-resolved traces. The Raman peaks in the spectra are annotated by the vertical grey lines at their respective spectral positions. Colored vertical bands in the FFT spectra are in one-to-one correspondence to their associated vibrational modes in the molecule indicated in their chemical structures in the left-panels. An acquisition time of one second was used for the time-resolved XPM traces, which corresponds to averaging over ~ $40 \times 10^3$ pump-probe delay scans with the photon-counting methodology, with one pump-probe delay scan requiring ~ 25 μs.

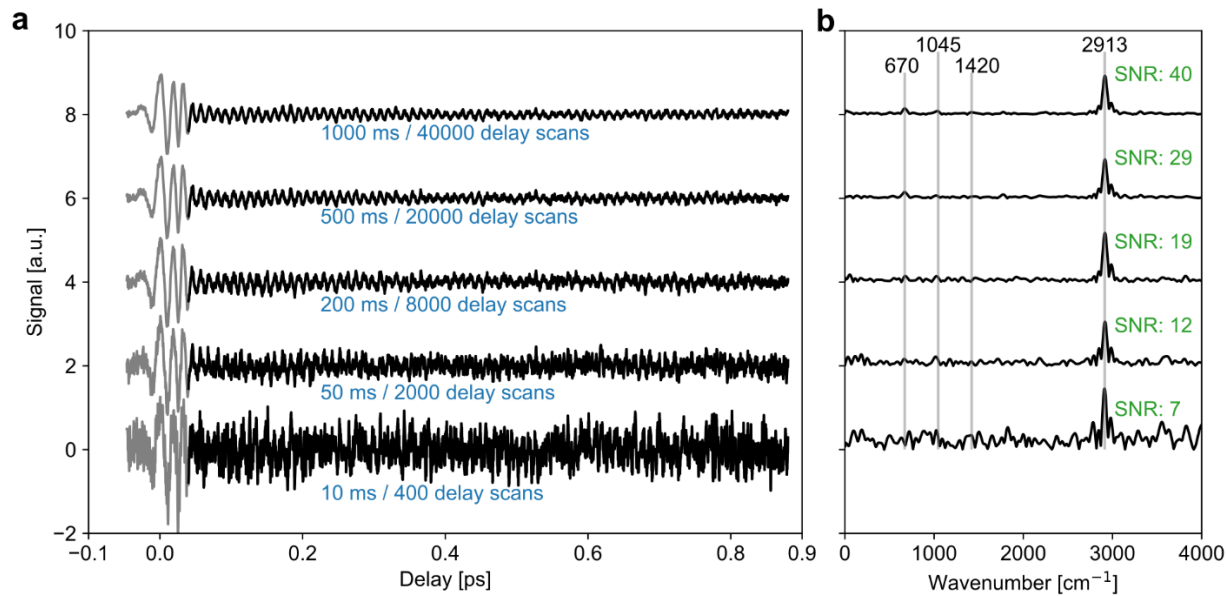

**Figure 3 | Photon-counting of the anti-Stokes emission in cross-phase modulation. a,** A series of time-resolved XPM traces measured in DMSO as a function of the delay between the pump and the probe pulses with varying averaging times. The traces are vertically shifted for clarity and the averaging times as well as the number of delay scans are annotated to each trace. **b,** Corresponding FFT spectra of the time-resolved XPM traces in **a**. The FFT spectra are horizontally aligned with the XPM traces in **a**. The Raman peaks in the spectra are annotated by the vertical grey lines at their respective spectral positions. Lower averaging of the pump-probe delay scans in the photon-counting exhibits spectra with dominant Raman peaks, nevertheless, the signal-to-noise of the time traces and the FFT spectra improve with higher averaging in the photon-counting. The spectra are vertically shifted for clarity and the single-to-noise ratios (SNR) are annotated to each spectrum.

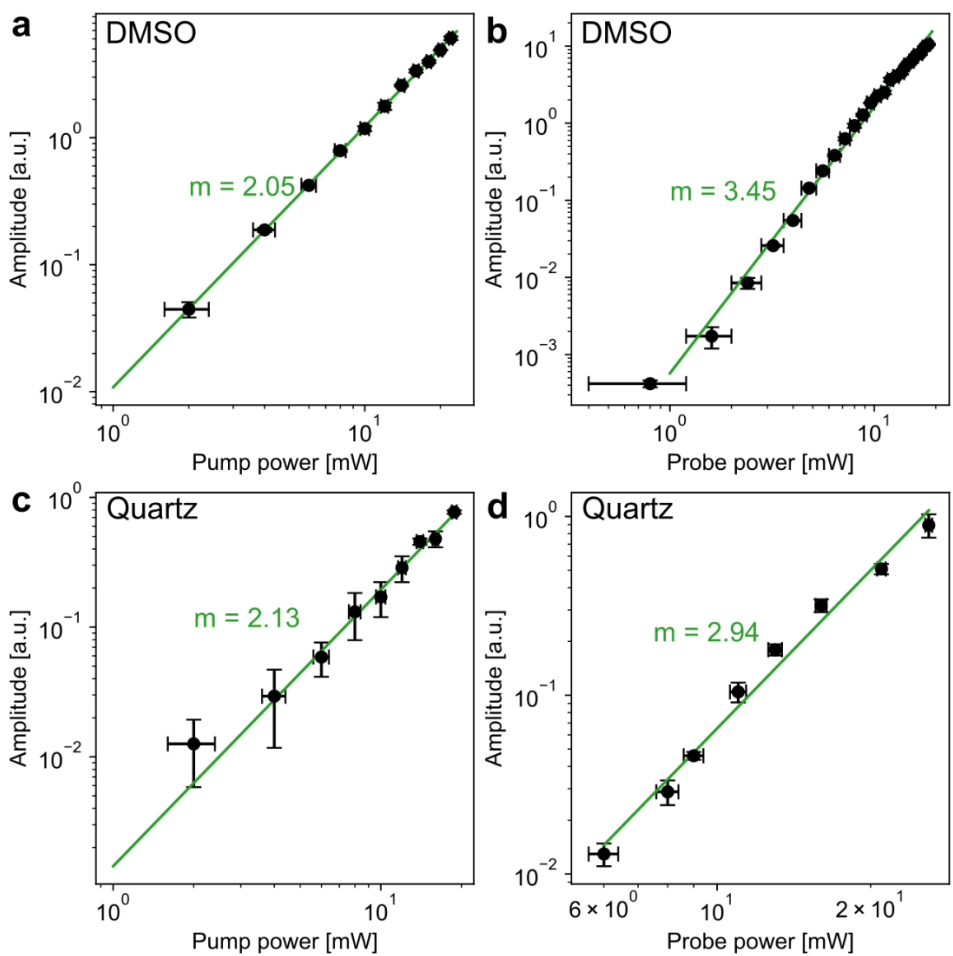

**Figure 4 | Evolution of the XPM signal with the fluence of the driving laser pulses. a, b,** Variation of the FFT amplitude of the molecular vibration at ~ 2913 cm⁻¹ retrieved from the time-resolved XPM traces measured in DMSO as a function of the increasing power of the pump and probe pulses, respectively. The power of the probe pulse and the pump pulse in **a** and **b** was kept constant at 29 mW and 19 mW, respectively. **c, d,** Variation of the FFT amplitude of the $A_{1b}$ mode (~ 462 cm⁻¹) of ~ 50 µm thick z-cut Quartz retrieved from the time-resolved XPM traces measured as a function of the increasing power of the pump and probe pulses, respectively. The power of the probe pulse and the pump pulse in **c** and **d** was kept constant at 29 mW and 19 mW, respectively. Green lines in **a-d** are the nonlinear fittings to the experimental data (black circles), with the order of nonlinearity (*m*) annotated to each fitting curve. Vertical error bars indicate the standard error in the maximum FFT amplitudes as retrieved from three consecutive measurements. Horizontal error bars denote the standard error in power measurements of the pump and the probe pulses.

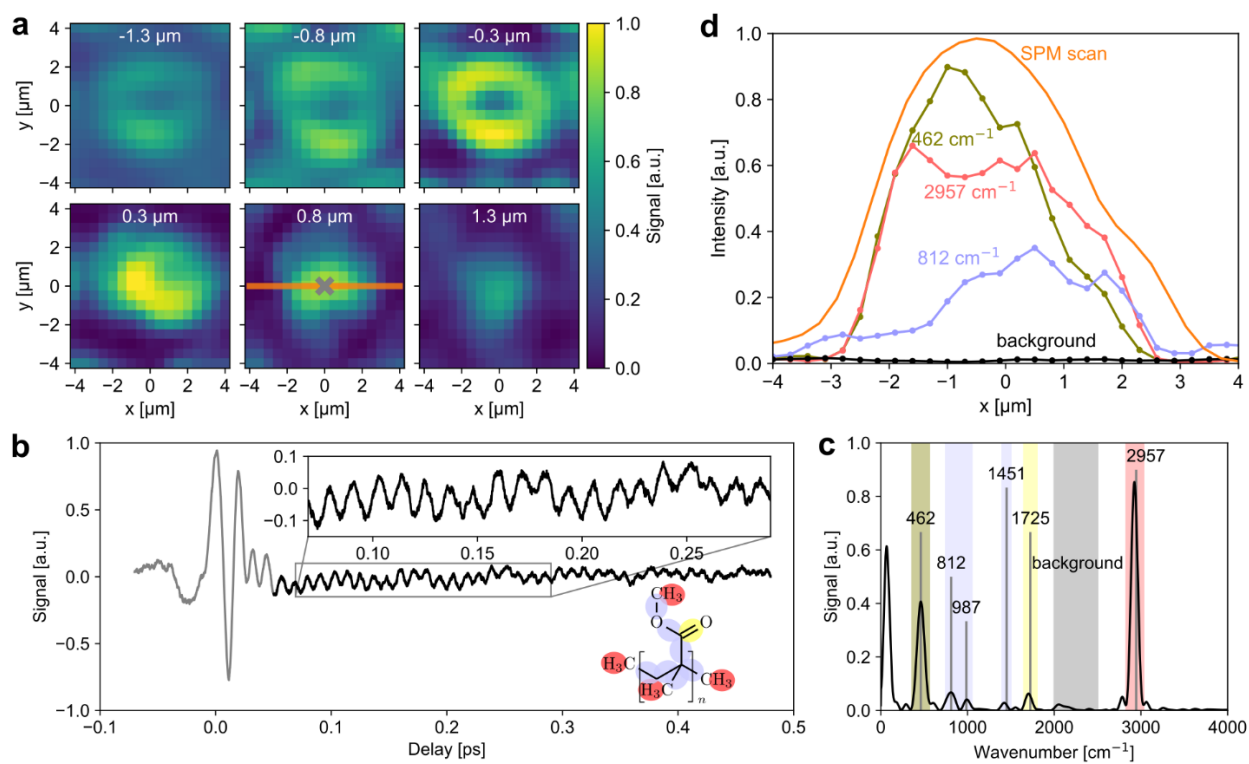

**Figure 5 | Hyperspectral chemical imaging. a.** A series of two-dimensional images of the SPM signal measured from a single PMMA bead of ~ 8 μm size at various positions of the bead along the focal axis, as annotated at the top of each image. The PMMA bead was drop casted onto a ~ 50 μm thick Quartz wafer. The anti-Stokes signal in SPM is generated by interaction of the sample only with the probe pulses (power ~ 29 mW) and was measured by the single-photon detector with an integration time of ~ 25 milliseconds. The PMMA bead was raster-scanned with a two-dimensional piezo stage across the fixed laser focal spot. **b,** Time-resolved XPM signal measured as a function of the delay between the pump and the probe pulses. The position of the laser spot on the bead is indicated by the grey colored cross in the second to last panel in **a**. An averaging time of 1 s in the pump-probe delay scans was used for the XPM signal in the photon-counting. The power of the pump and probe pulses were set at ~ 19 mW and ~ 29 mW, respectively. Inset shows the zoom-in of the XPM trace in a narrow time window. **c,** FFT spectrum retrieved from the time-resolved XPM signal shown in **b**, with the colored vertical bands indicating the spectral positions of the annotated Raman peaks. The spectral positions of the colored vertical bands correspond one-to-one to the associated vibrational mode of PMMA, whose chemical structure is shown in **b. d,** Spatial variation of the FFT amplitude of the Raman peaks as retrieved from the time-resolved XPM signal measured at 30 equidistant positions across the PMMA bead, as annotated by the orange-line in the second last panel in **a**. The spatial variations of the Raman peaks are color-coded one-to-one with the colored vertical bands in **c.**

The orange curve shows the spatial variation of the SPM signal measured only with the probe pulses at the identical positions of the laser focal spot on the bead as for the XPM signal.